\documentclass{aastex631}

\graphicspath{{./}{figures/}}
\shortauthors{Dalal et al.}
\usepackage{xcolor}
\usepackage{soul}

\begin{document}

\title{Suprathermal population associated with stream interaction regions observed by STEREO-A: New insights } 

\author{Bijoy Dalal}
\affiliation{Physical Research Laboratory, Ahmedabad - 380009, India}
\affiliation{Indian Institute of Technology Gandhinagar, Gandhinagar - 382055, India}

\author{Dibyendu Chakrabarty}
\affiliation{Physical Research Laboratory, Ahmedabad - 380009, India}

\author{Nandita Srivastava}
\affiliation{Udaipur Solar Observatory, Physical Research Laboratory, 
Udaipur - 313001, India}

\author{Aveek Sarkar}
\affiliation{Physical Research Laboratory, Ahmedabad - 380009, India}

\begin{abstract}
Stream interaction regions (SIRs) are often thought to be responsible for the generation of suprathermal population in the interplanetary medium. Despite the source being same, wide variations in spectral indices of suprathermal populations are observed at 1 au during SIRs. This poses significant uncertainty in understanding the generation of suprathermal ion populations by SIRs and indicates interplay of multiple source mechanisms. In the present work, by analyzing variations in suprathermal $^4$He, O, and Fe for 20 SIR events recorded by STEREO-A during 2007 – 2014, we find that the spectral indices of these elements vary in the range of 2.06 – 4.08, 1.85 - 4.56, and 2.11 – 4.04 respectively  for 19 events. However, in one special case, all the three suprathermal elements show nearly identical ($\sim$1.5) spectral indices. We offer possible mechanisms, which can cause significant variations in the spectral indices of suprathermal particles. More importantly, we show the possible role of merging and/or contraction of small-scale magnetic islands near 1 au in producing nearly identical spectral indices for three different elements with different first ionization potential and mass-to-charge ratio. The occurrence of these magnetic islands near 1 au also supports the minimal modulation in spectral indices of these particles. The role of a possible solar flare in generating these magnetic islands near the heliospheric current sheet is also suggested. 
\end{abstract}


\section{Introduction} \label{sec:intro}
Suprathermal particles with energies in the range 0.01 – 1.0 MeV per nucleon (hence, MeV n$^{-1}$), often characterized by power law distributions, are ubiquitous in the heliosphere \citep{Gloeckler_2003}. The power law index (spectral index) was, in general, suggested to be - 1.5 due to ``stochastic acceleration in compressional turbulences” during ``quiet” times if differential directional flux is plotted against energy of the particles (\citealp{Gloeckler_2003, Fisk_and_Gloeckler_2006, Fisk_and_Gloeckler_2007, Fisk_and_Gloeckler_2008, Fisk_and_Gloeckler_2014} etc.) Generation of suprathermal particles in the interplanetary (IP) medium is one of the most debatable topics for several decades. Stream interaction regions (SIRs) are considered as major source of suprathermal and energetic particles in the IP medium (\citealp{Tsurutani_et_al_1982, Mason_et_al_2012, Richardson_2018, Allen_et_al_2019, Allen_et_al_2021} etc.). SIRs form when faster solar wind streams from coronal holes interact with preceding slower streams originating from coronal streamers \citep{Belcher_and_Davis_1971, Richardson_2018}. Sometimes, coronal holes and fast solar wind emanating from those persist for several solar rotations (Carrington rotations $\sim$ 27 days). This causes the compressed interaction regions to corotate with the Sun for more than one solar rotation. These structures are then called corotating interaction regions (CIRs) \citep{Gosling_et_al_1981}. Spectral indices of SIR/CIR associated suprathermal $^4$He calculated over energy 0.1 – 1 MeV n$^{-1}$ observed at 1 au may vary in the range 1.3 – 5 \citep{Mason_et_al_2008a, Allen_et_al_2021}. \cite{Mason_et_al_2008a} found that spectral indices of CIR associated $^4$He and O in the energy range 0.16 – 0.91 MeV n$^{-1}$ are correlated and similar. However, slopes of Fe spectra are found to be different from those of $^4$He and O. This raises the question whether acceleration of SIR/CIR-associated suprathermal ions depend on first ionization potential (FIP) and mass-to-charge ratio (m/q) of species. In a recent paper, \cite{Dalal_et_al_2022} have shown that modulation of quiet time suprathermal particles in the IP medium depends on FIP and m/q of elements. In particular, elements with contrasting FIP and m/q (like $^4$He and Fe) behave differently during quiet times in solar cycle 24 as compared with cycle 23 while, species with similar FIP and m/q (like C and O) are modulated similarly in both the solar cycles. 
\begin{figure}[ht!]
\plotone{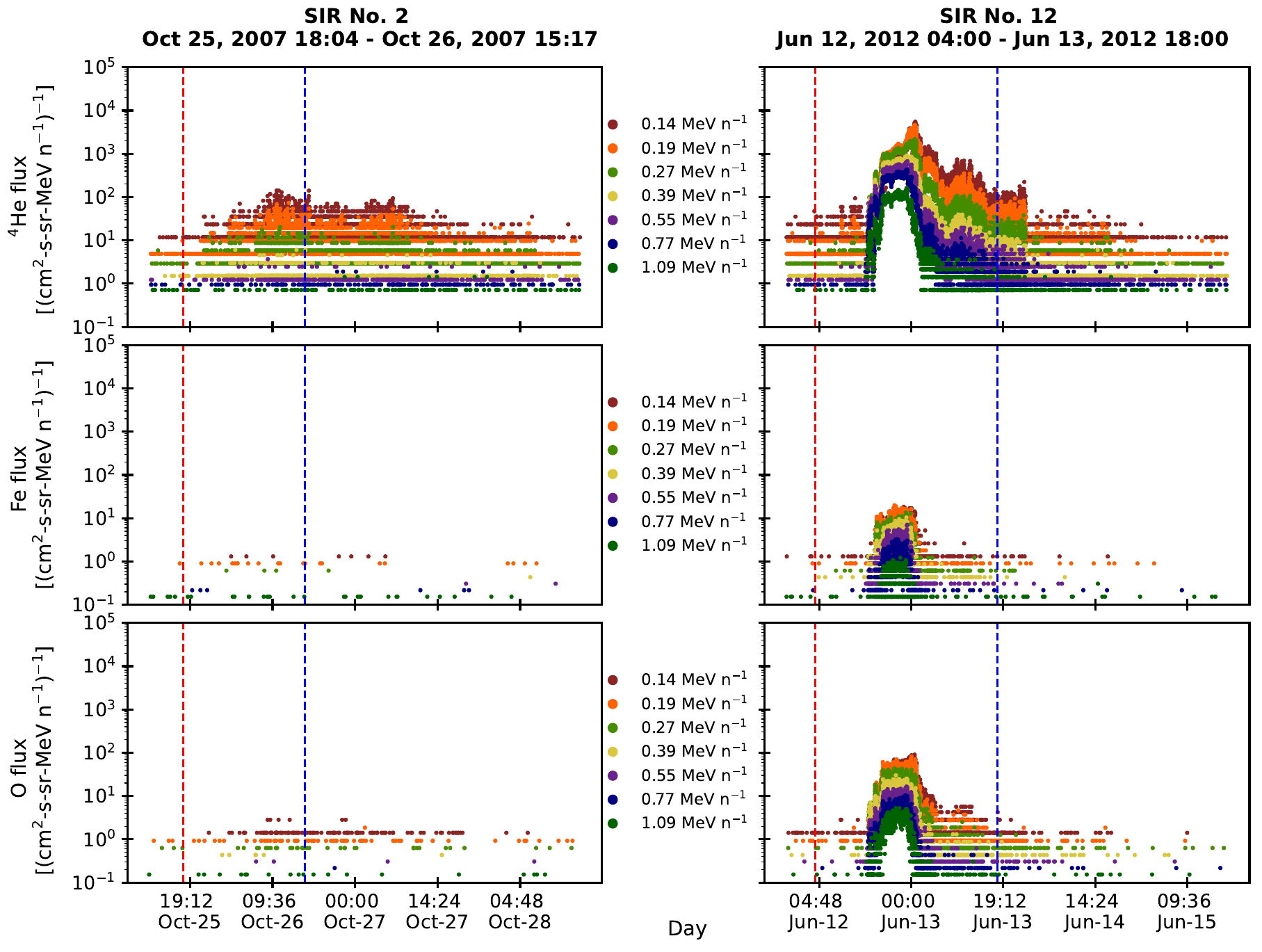}
\caption{Temporal variations of suprathermal $^4$He (top row), Fe (middle row), and O (bottom row) fluxes at different energy channels associated with two typical SIR events: SIR No. 2 (left) and SIR No. 12 (right). Table \ref{tab:table1} provides the details on the date and time of arrival of these SIRs at the location of STEREO-A. The start and end times of these events (determined by the characteristic changes in solar wind parameters associated with SIR/CIR) are shown by red and blue vertical dashed lines respectively. The intervals of these events are written in the titles of the left and right column of the figure. Energy channels are mentioned in between the columns.\label{fig:fig1}}
\end{figure}

One of the existing theories related to particle acceleration in SIR/CIRs is that suprathermal and energetic particles are generated in forward and reverse shocks associated with leading and trailing edges of SIRs respectively \citep{Fisk_and_Lee_1980}. However, these shocks may or may not form at 1 au \citep{Richardson_2018} and suprathermal particles may intensify during the passage of shock-less SIR/CIR events as well (\citealp{Giacalone_et_al_2002, Mason_et_al_2008a, Buvcik_et_al_2009, Ebert_et_al_2012, Filwett_et_al_2019, Allen_et_al_2019, Allen_et_al_2021} etc.). Note, the ``shock-less” events do not exclude presence of shock accelerated particles at the satellite location. As discussed by \cite{Fisk_and_Lee_1980}, contribution from suprathermal particles generated by reverse shock beyond 1 au may cause enhancements in suprathermal fluxes at 1 au. 

In the present work, we select the events wherein the signatures of shocks are not identified at the satellite location. It is also to be noted that it is not only shocks but the gradual compression regions are also suggested \citep{Giacalone_et_al_2002} to play an important role in the generation of suprathermal population in the interplanetary (IP) medium. While the role of shocks in particle acceleration is well accepted, the role of gradual compression regions in generating suprathermal pool has not received enough attention. Further, if well-defined shock structures are detected by the spacecraft, it may dominate the suprathermal flux enhancements at the satellite location. Therefore, in order to ensure that we do not disregard the contribution of suprathermal population generated from other processes like the gradual compression regions, the so-called “shock-less” events are chosen. Having selected these events, we focus on the evaluation of the spectra of elements having different FIP and m/q. In this work, therefore, we estimate spectral indices of $^4$He, O, and Fe with energies 0.14 – 1.09 MeV n$^{-1}$ associated with twenty shock-less SIR events observed by Solar Terrestrial Relations Observatory – Ahead (STEREO-A). We note significant variations in spectral indices of across the SIR events and elements similar to “quiet-time” conditions (e.g. \citealp{Dalal_et_al_2022}). The possible physical processes that might contribute to this spectral index variability during SIR events are discussed. Interestingly, we also find one special SIR event when these three elements exhibit nearly identical spectral indices. We investigate this special SIR event in detail and argue that the enhancement in the suprathermal population during this event is a consequence of merging and/or contraction of small-scale magnetic island near the spacecraft. The following section discusses about the data used in this work. In subsequent section, we present our results. ``Discussions and conclusions” follows the ``Results” section. We summarize the results in the ``Summary” section.

\begin{figure}[ht!]
\plotone{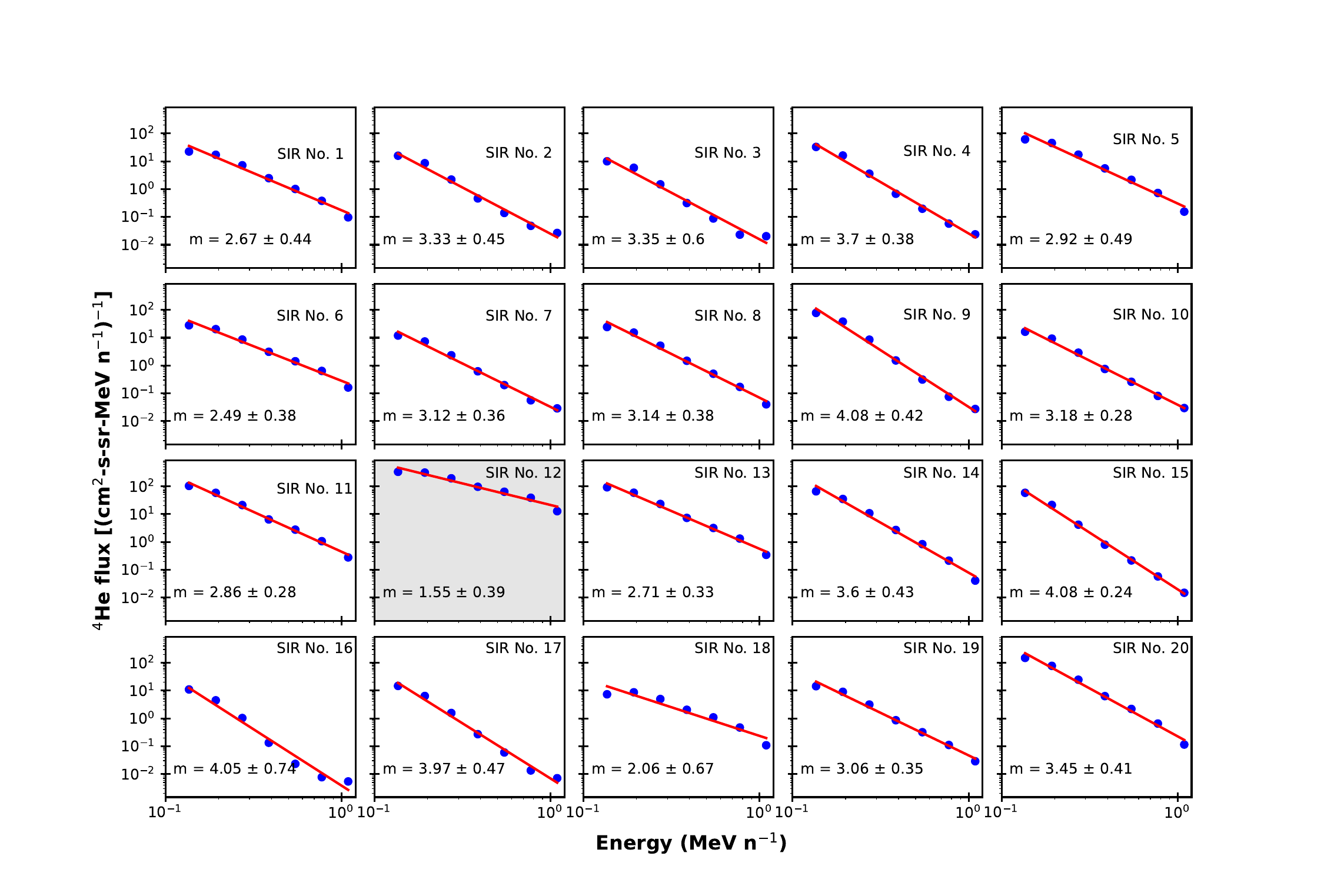}
\caption{Differential directional flux vs energy spectra of $^4$He during flux enhancements associated with SIR No. 1 – 20. Each subplot corresponds to an SIR event written at the right upper corner of the subplot. Spectral indices (m) along with margin of errors (MoEs) within 95$\%$ of confidence bounds as errors are written in the bottom left of each subplot. It is noted that the spectral indices of $^4$He vary from 1.55 to 4.08. SIR No. 12 is marked with grey background, as the spectral index for this event is 1.55 and is distinctively less as compared to the rest of the events.\label{fig:fig2}}
\end{figure}
\begin{figure}[ht!]
\plotone{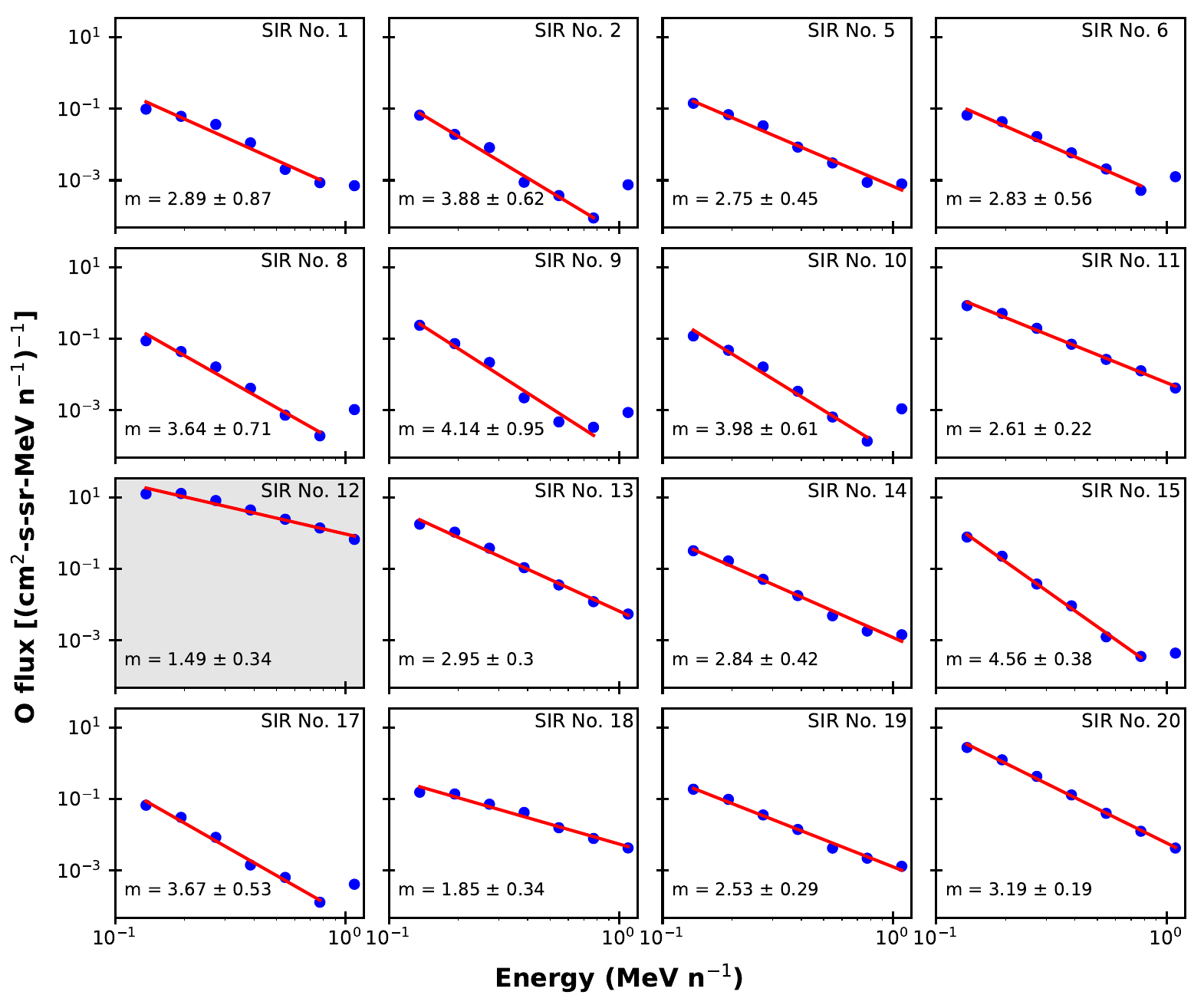}
\caption{Spectra of O during the enhancements in SIR events mentioned in the upper right corner of each subplot. The spectral indices (m) and MoEs are written in the bottom left of all the boxes. The spectral indices here vary from 1.49 to 4.56. Similar to Figure \ref{fig:fig2}, SIR No. 12 is marked with grey background, as spectral index is 1.49 in this event, which distinctively less compared to the rest of the events.\label{fig:fig3}}
\end{figure}
\section{Data used}
The SIR events used in this work are observed by Solar Terrestrial Relations Observatory – Ahead (STEREO-A) during 2007 – 2014 and are available at \url{https://stereo-dev.epss.ucla.edu/media/SIRs.pdf}. The start, end, and approximate passage times of the stream interface structure are determined based on criteria discussed in \cite{Jian_et_al_2006, Jian_et_al_2013, Jian_et_al_2019}. Suprathermal $^4$He, O, and Fe fluxes with energies $< $ 1 MeV n$^{-1}$ obtained from Suprathermal Ion Telescope (SIT) \citep{Mason_et_al_2008b} during and close to these events are used. SIT is a time of flight mass spectrometer with a mass resolution ($\sigma_m$/m) of 0.1 in the 0.1 – 1 MeV n$^{-1}$ energy range. This mass-resolution is enough to distinguish $^3$He and $^4$He in the said energy range (see \citealp{Mason_et_al_2008b} for details). Suprathermal particle flux data for $^4$He, O, and Fe can be found at \url{https://cdaweb.gsfc.nasa.gov/index.html}. These elements have different FIP and m/q. It is also checked if any IP shock is observed by STEREO-A at the time of the event.  List of IP shocks observed by STEREO-A is available at \url{http://www.ipshocks.fi/database}. Solar wind proton density and bulk speed data are obtained from the Plasma and Suprathermal Ion Composition (PLASTIC) instrument \citep{Galvin_et_al_2008} on board STEREO-A. The components of interplanetary magnetic field (IMF) in Radial-Tangential-Normal (RTN) coordinate system are measured by the fluxgate magnetometer \citep{Acuna_et_al_2008}, a subsystem of  the In-situ Measurements of Particles and CME Transients (IMPACT) investigation \citep{Luhmann_et_al_2008} on board STEREO-A. These data are also available at \url{https://cdaweb.gsfc.nasa.gov/index.html}. The event-selection criteria adopted in this work are discussed in the next section.  
\section{Selection of events}
Among all the SIR events listed in the above link (\url{https://stereo-dev.epss.ucla.edu/media/SIRs.pdf}), we have selected twenty SIR events during 2007 – 2014 when there are noticeable enhancements in the suprathermal $^4$He. These events are denoted by SIR No. 1 – 20 and listed in Table \ref{tab:table1}. As discussed earlier, these are ``shock-less” SIR events. It is noted that not every SIR event is associated with enhancements in the O and Fe fluxes. Sixteen events are observed when O fluxes show small or moderate enhancements and nine events are seen when Fe fluxes show enhancements. Figure \ref{fig:fig1} shows the typical temporal variations of $^4$He, Fe, and O fluxes at different energy channels corresponding to SIR No. 2 and SIR No. 12. One can easily see that $^4$He fluxes have increased noticeably above the background (level before and after the enhancement) in connection with SIR No. 2. The enhancement in Fe fluxes is negligible and averaged O fluxes have negligible increment. On the other hand, large enhancements in all the three elements are observed in association with SIR No. 12. The second and third columns of Table \ref{tab:table1} enlist the start and end times of all the twenty SIRs events observed by STEREO-A.

\begin{deluxetable*}{cccccc}[ht!]
\tablecolumns{6}
\tablenum{1}
\tablecaption{List of SIR-events and spectral indices of $^4$He, O, and Fe corresponding to those events. \label{tab:table1}}
\tablewidth{0pt}
\tablehead{
\colhead{SIR No.} &
\colhead{Start time} &
\colhead{End time} & {} & \colhead{Spectral index} & {}\\
\cline{4-6}
  &  &  & $^4$He & O & Fe
}
\startdata
1  & Sep 21, 2007 13:35 & Sep 23, 2007 11:32 & 2.67$\pm$0.44 & 2.89$\pm$0.87             & {}           \\
2  & Oct 25, 2007 18:04 & Oct 26, 2007 15:17 & 3.33$\pm$0.45 & 3.88$\pm$0.62            & {}           \\
3  & Nov 13, 2007 18:00 & Nov 16, 2007 03:30 & 3.35$\pm$0.60 & {}            & {}           \\
4  & Dec 11, 2007 18:00 & Dec 13, 2007 04:10 & 3.70$\pm$0.38 & {}            & {}           \\
5  & Feb 11, 2008 00:00 & Feb 13, 2008 04:40 & 2.92$\pm$0.49 & 2.75$\pm$0.45 & 2.59$\pm$0.69          \\
6  & Feb 29, 2008 16:00 & Mar 03, 2008 03:15 & 2.49$\pm$0.38 & 2.83$\pm$0.56            & {}           \\
7  & Jun 16, 2008 06:00 & Jun 17, 2008 18:36 & 3.12$\pm$0.36 & {}            & {}           \\
8  & aug 07, 2008 18:10 & aug 12, 2008 00:00 & 3.14$\pm$0.38 & 3.64$\pm$0.71            & {}           \\
9  & May 31, 2009 08:13 & Jun 02, 2009 16:00 & 4.08$\pm$0.42 & 4.14$\pm$0.95 & {}           \\
10 & Jan 12, 2011 12:03 & Jan 14, 2011 08:00 & 3.18$\pm$0.28 & 3.98$\pm$0.61            & 4.04$\pm$1.57           \\
11 & Nov 12, 2011 02:00 & Nov 14, 2011 18:00 & 2.86$\pm$0.28 & 2.61$\pm$0.22 & 2.09$\pm$0.27    \\
12 & Jun 12, 2012 04:00 & Jun 13, 2012 18:00 & 1.55$\pm$0.39 & 1.49$\pm$0.34 & 1.46$\pm$0.60 \\
13 & aug 21, 2012 20:00 & aug 23, 2012 11:10 & 2.71$\pm$0.33 & 2.95$\pm$0.30 & 3.09$\pm$0.46 \\
14 & Apr 18, 2013 16:00 & Apr 19, 2013 14:19 & 3.60$\pm$0.43 & 2.84$\pm$0.42 & 3.09$\pm$0.46 \\
15 & Jul 17, 2013 00:00 & Jul 22, 2013 16:00 & 4.08$\pm$0.24 & 4.56$\pm$0.38 & {}           \\
16 & Sep 24, 2013 08:00 & Sep 25, 2013 12:00 & 4.05$\pm$0.74 & {}            & {}           \\
17 & Oct 24, 2013 10:00 & Oct 26, 2013 02:25 & 3.97$\pm$0.47 & 3.67$\pm$0.53            & {}           \\
18 & Apr 28, 2014 16:47 & Apr 30, 2014 19:05 & 2.06$\pm$0.67 & 1.85$\pm$0.34 & 2.11$\pm$0.59\\
19 & May 06, 2014 02:37 & May 07, 2014 02:00 & 3.06$\pm$0.35 & 2.53$\pm$0.29            & 2.82$\pm$0.51\\
20 & Jun 03, 2014 07:27 & Jun 05, 2014 00:00 & 3.45$\pm$0.41 & 3.19$\pm$0.19 & 2.72$\pm$0.76\\
\enddata
\end{deluxetable*}

\section{Results: Spectral analysis}
As is known, suprathermal ions ($<$ 1 MeV n$^{-1}$) at 1 au exhibit power law in the differential flux vs energy spectra. In the present investigation, we use spectral indices associated with these power laws as representative of the processes that generate the suprathermal particles corresponding to the SIR events. The spectra of $^4$He during the enhanced flux level corresponding to SIR No. 1 – 20 are shown in Figure \ref{fig:fig2}. The spectral indices and associated margin of errors (MoEs) within 95$\%$ of confidence bounds are listed in the fourth column of Table \ref{tab:table1}. \cite{Dalal_et_al_2022} has used MoEs as error bars in estimated spectral indices of quiet suprathermal particles rigorously and for detailed discussion on this, one may refer to \cite{Agresti_and_Coull_1998} as well as \cite{Dalal_et_al_2022} One can see from Figure \ref{fig:fig2} and Table \ref{fig:fig1} that spectral indices of suprathermal $^4$He lie between 1.55 and 4.08 for these events. 

\begin{figure}[h!]
\plotone{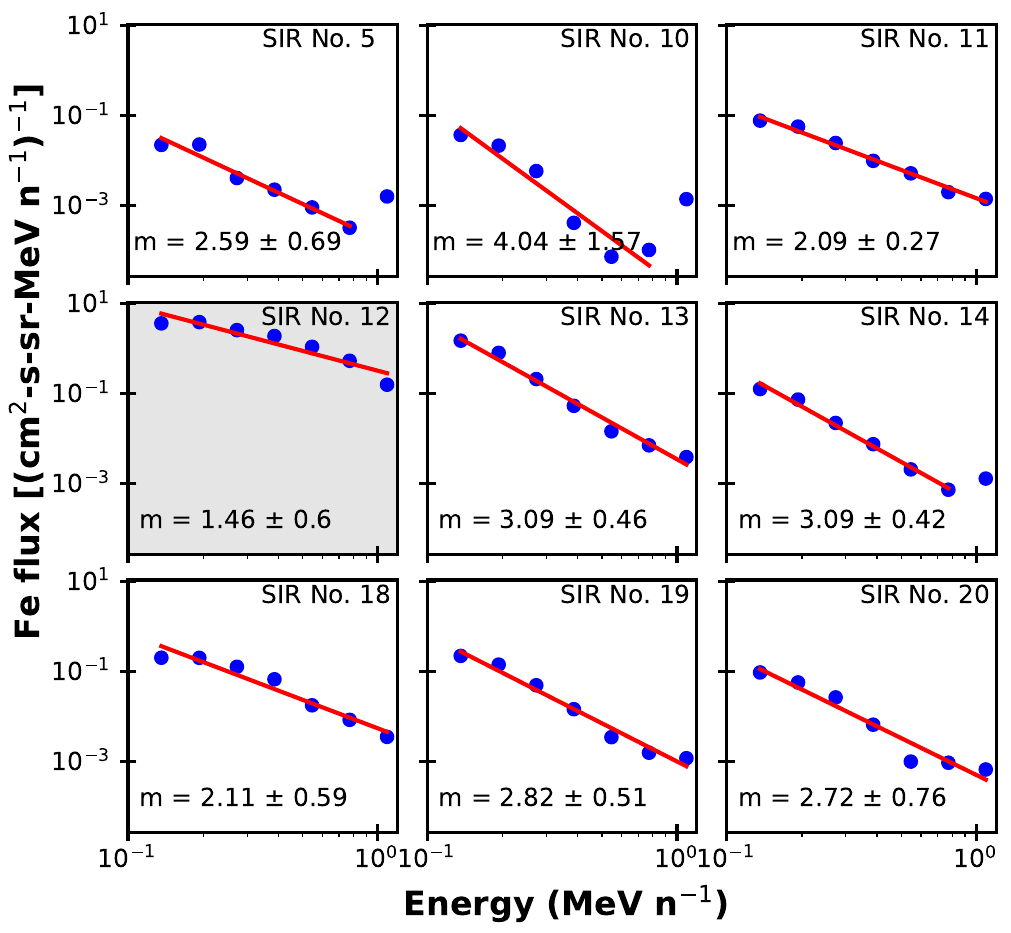}
\caption{Spectra of Fe during the enhancements in SIR events mentioned in the upper right corner of each subplot. The spectral indices (m) and MoEs are written in the bottom left of all the boxes. In case of Fe, the spectral indices vary from 1.46 to 4.04. Similar to Figure \ref{fig:fig2} and \ref{fig:fig3}, SIR No. 12 is also marked with grey background to highlight the distinctively less spectral index (1.46) as compared to the rest of the events.\label{fig:fig4}}
\end{figure}

Figure \ref{fig:fig3} and Figure \ref{fig:fig4} show respectively the spectra of O and Fe corresponding to those SIR events when we see enhancements in these elements. Their spectral indices are also listed in Table \ref{tab:table1}. It is noted that O and Fe spectra are sometimes best fitted up to 0.77 MeV n$^{-1}$ as shown in Figure \ref{fig:fig3} and Figure \ref{fig:fig4}. Spectral indices of O and Fe vary from 1.49 to 4.56 and from 1.46 to 4.04 respectively. Although the spectral indices for $^4$He, O, and Fe are, in general, different corresponding to other SIR events, these are almost equal in case of SIR No. 12 and are very close to 1.5 (1.55, 1.49, and 1.46 respectively). This is consistent with the fact that the nature of flux variations of all the three elements corresponding to SIR No. 12 (shown in right panel of Figure \ref{fig:fig1}) are different (impulsive increase followed by gradual decay in the flux) from the nature of flux variations during other SIR events.

\begin{figure}[h!]
\plotone{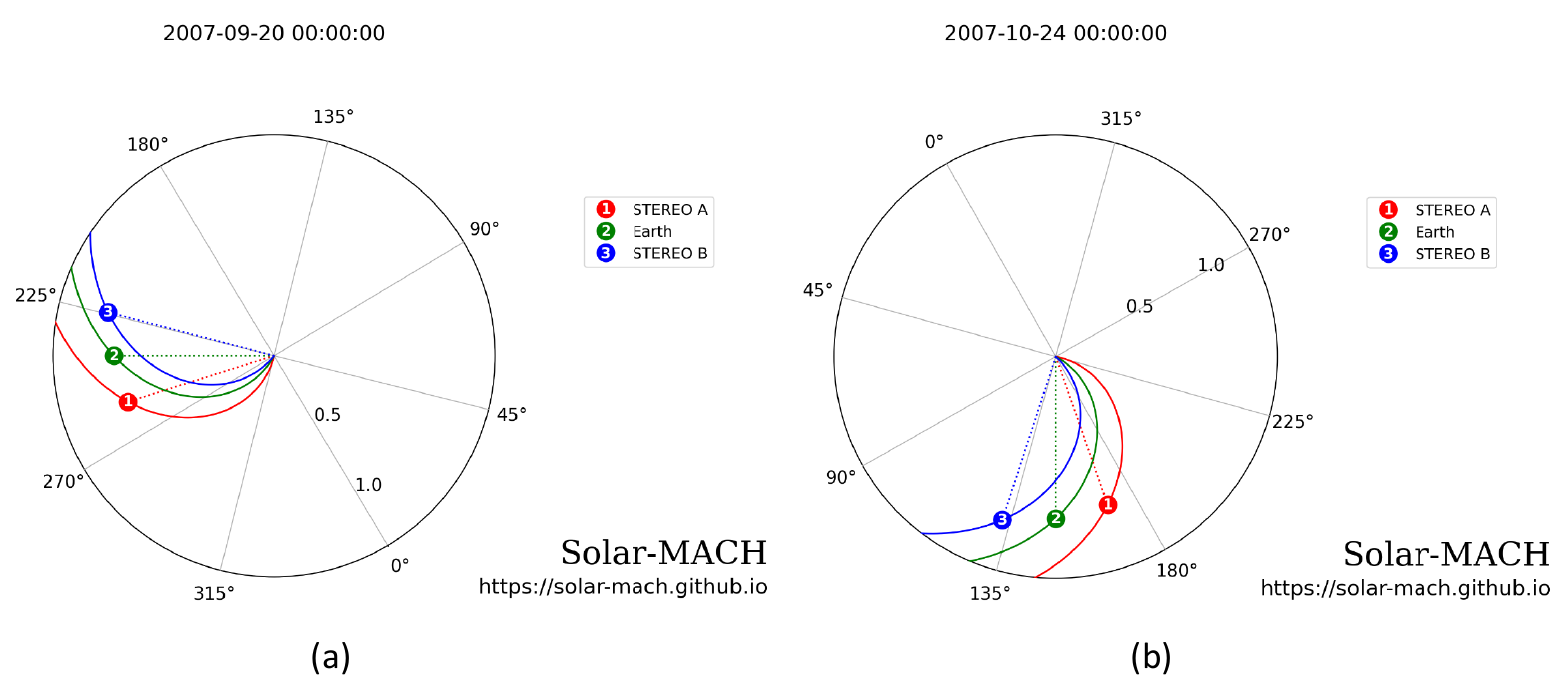}
\caption{Relative positions of (1) STEREO-A (red), (2) the Earth (green), and (3) STEREO-B (blue) on (a) Sep 20, 2007 at 00:00 hour and (b) Oct 24, 2007 at 00:00 hour respectively. The colored solid lines are Parker spirals connecting these objects and are calculated based on solar wind speed of 400 km/s. This figure is available at \url{https://solar-mach.streamlit.app/?embedded=true}.\label{fig:fig5}}
\end{figure}

\begin{figure}[h!]
\plotone{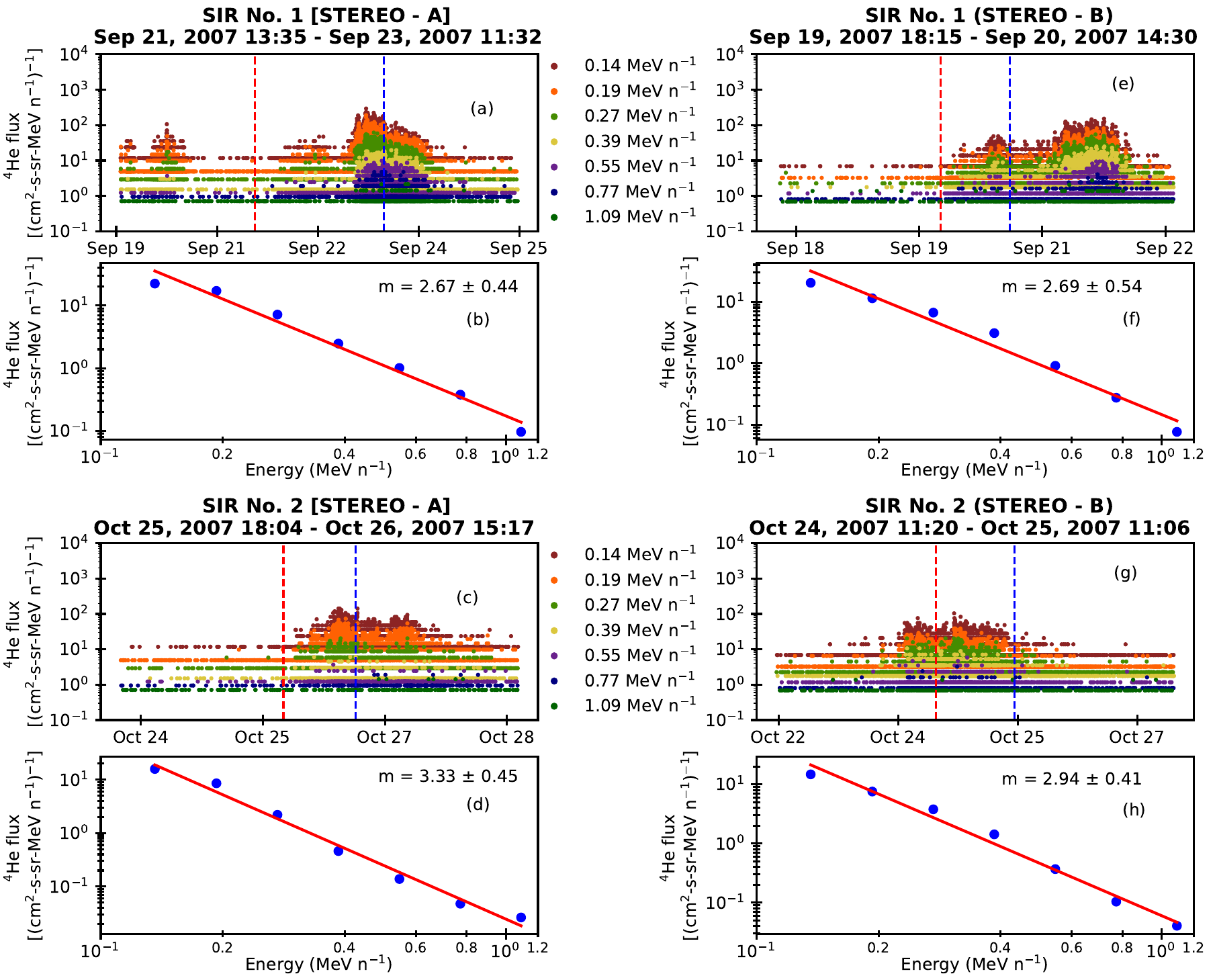}
\caption{Temporal variations of $^4$He fluxes at different energies (mentioned in between) measured by [(a), (c)] STEREO-A and [(e), (g)] STEREO-B corresponding to two pairs of SIR events (SIR No. 1 (STEREO-A, STEREO-B) and SIR No.2 (STEREO-A, STEREO-B)) observed by these two spacecraft when those were very close to each other. Spectra of $^4$He during the enhanced periods associated with SIR No. 1 (STEREO-A, STEREO-B) and SIR No. 2 (STEREO-A, STEREO-B) are shown in panel (b, f) and (d, h) respectively. Spectral indices (m) with MoEs are also mentioned.\label{fig:fig6}}
\end{figure}

In addition to twenty SIR events observed by STEREO-A, we also calculate spectral indices for suprathermal $^4$He associated with two SIR events observed by STEREO-B. Interestingly, SIR No. 1 and SIR No. 2 (see Table \ref{tab:table1}) had been observed by STEREO-B before they were observed by STEREO-A (a list of SIR events observed by both the STEREO spacecraft is available at \url{https://stereo-ssc.nascom.nasa.gov/data/ins_data/impact/level3/}). Figure \ref{fig:fig4} (a) and Figure \ref{fig:fig5} (b) show the positions of STEREO-A and STEREO-B with respect to the Earth on Sep 20, 2007 and Oct 24, 2007 respectively. Figure \ref{fig:fig6} (a) and (e) show $^4$He flux variations corresponding to SIR No. 1 (STEREO-A) (started on Sep 21, 2007 13:37) and SIR No. 1 (STEREO-B) (started on Sep 19, 2007 18:15) respectively. The corresponding $^4$He spectra are shown in Figure \ref{fig:fig6} (b) and (f) respectively. The spectral indices for $^4$He observed by STEREO-A and STEREO-B are 2.67 and 2.69 respectively. Figure \ref{fig:fig6} (c) and (g) show $^4$He flux variations during another pair of SIR events observed by STEREO-A and STEREO-B. This time, spectral indices are 3.33 and 2.94 respectively as observed by STEREO-A and STEREO-B. This exercise shows that spectral index of SIR-associated $^4$He observed by closely spaced spacecraft may also vary. We discuss all these aspects in the next section.

\section{Discussions and conclusions}
In this work, it is shown that the fluxes of all suprathermal elements (like $^4$He, O, and Fe) having similar energies per nucleon may not get enhanced during an SIR event. One of the probable reasons for this could be the lack of seed populations in the IP medium that are subjected to acceleration (or, deceleration) and eventually become part of the suprathermal pool of particles.  Another possibility may be linked to the efficacy of the generation mechanism of suprathermal ions in the IP medium. \cite{Ebert_et_al_2012} showed that peak intensity of $<$ 0.8 MeV n$^{-1}$ $^4$He population is correlated with the local magnetic compression ratios (M, defined as ratio of downstream and upstream magnetic fields) at 1 au even in the absence of reverse shocks. The correlation between local compression ratio strength and $\sim$ 1 MeV n$^{-1}$ proton intensities at 5 au was established by \cite{Desai_et_al_1998}. Therefore, it appears that local compression ratio strength plays an important role in shaping the intensity variations of $<$ 1 MeV n$^{-1}$ suprathermal particles. However, the critical value of compression ratio strength required to raise suprathermal fluxes above background level is not clear from those studies.  

It is brought out from the spectral analysis of $^4$He, O, and Fe fluxes that there are wide variations in spectral indices of these elements across different shock-less SIR events. The spectral indices lie in the range 1.55 - 4.08, 1.49 - 4.56 and from 1.46 - 4.04 respectively for $^4$He, O, and Fe. \cite{Mason_et_al_2008a} and \cite{Allen_et_al_2021} also reported spectral indices of similar range in the energy range 0.16 – 0.91 MeV n$^{-1}$ in case of CIR associated $^4$He measured by the Ultra-Low Energy Isotope Spectrometer (ULEIS) on board the Advanced Composition Explorer (ACE) and the Suprathermal Ion Spectrograph (SIS) on board the Solar Orbiter. Some of the SIR/CIR events investigated in \cite{Mason_et_al_2008a} are associated with shocks. Even if there are no well-formed shocks at 1 au, acceleration of particles is still possible in gradual compression regions associated with SIR/CIRs \citep{Giacalone_et_al_2002}. \cite{Filwett_et_al_2019} also reported variations in spectral indices of suprathermal particles in different SIR events. It is worth noting that suprathermal heavy ions (like O, Fe) associated with IP shocks also show spectral indices varying from 1.00 to 4.00 \citep{Desai_et_al_2004} in the energy range of 0.1 - 0.5 MeV n$^{-1}$. Therefore, wide variations in spectral indices of suprathermal particles below 1 MeV n$^{-1}$ seems to be a common feature in case of stochastically accelerated particles.  

In order to check the spatial variation of spectral index of suprathermal particles associated with SIR events, we compare spectra of $^4$He corresponding to two SIR events (SIR No. 1 and 2 in Table \ref{tab:table1} observed at first by STEREO-B and then by STEREO-A when these satellites were very close (around 84 Gm) to each other. Despite proximity of STEREO-A and STEREO-B, these satellites recorded differences in solar wind parameters (like magnetic field, maximum proton density, extreme solar wind speeds across the SIRs etc.) in both the events. Interestingly, we find that in the first SIR event observed at first by STEREO-B and then by STEREO-A, the spectral indices of $^4$He are almost equal (STEREO-A: 2.67 and STEREO-B: 2.69). On the other hand, spectral indices for $^4$He observed by STEREO-B and STEREO-A differ (STEREO-A: 3.33 and STEREO-B: 2.94) in case of the second SIR event. On further investigation, we find that the second SIR event observed by STEREO-B (Figure \ref{fig:fig6}) followed an interplanetary coronal mass ejection (ICME) (list of ICMEs can be found at \url{https://stereo-ssc.nascom.nasa.gov/data/ins_data/impact/level3/}), while STEREO-A did not observe any ICME. This suggests that spectral index of suprathermal particles associated with SIRs may differ if the suprathermal particle population contains particles from more than one source. This proposition gets credence from the work of \cite{Schwadron_et_al_2010} wherein the authors argue that time-averaged spectra of particles are essentially superposition of many distribution functions related to different solar wind conditions like temperature, density etc. In other words, since SIRs acquire remnant particles from different events (like particles from an ICME in the second SIR event observed by STEREO-B) from different parts of the IP medium, during the course of their evolution, it is possible that many processes are at play in modulating suprathermal particle spectra. 

It is to be noted from Table \ref{tab:table1} that we see differences in spectral indices for different elements corresponding to 19 SIR events, except for SIR No. 12. Non-uniform spectral indices across different elements corresponding to particular SIR event have been reported earlier \citep{Filwett_et_al_2019}. \cite{Mason_et_al_2008a} reported that $^4$He and O spectral indices in the energy range 0.16 – 0.91 MeV n$^{-1}$ are extremely well correlated and average spectral index of O is 1.03 times that of $^4$He. However, this is not the case in our investigation. We find that spectra of $^4$He are harder in 50 $\%$ of the SIR events when we see enhancements in O fluxes (see column 4 and 5 of Table \ref{tab:table1}). Spectral index of Fe also does not match with spectral index of other elements. In terms of FIP, O ($\sim$13.6 eV) lies in between Fe ($\sim$7.9 eV) and $^4$He ($\sim$24.6 eV). These elements differ in m/q also. Therefore, it appears that the source and generation mechanisms involved in producing suprathermal particles associated with most of the shock-less SIR events at 1 au are possibly dependent on FIP and m/q values. 

\begin{figure}[h!]
\plotone{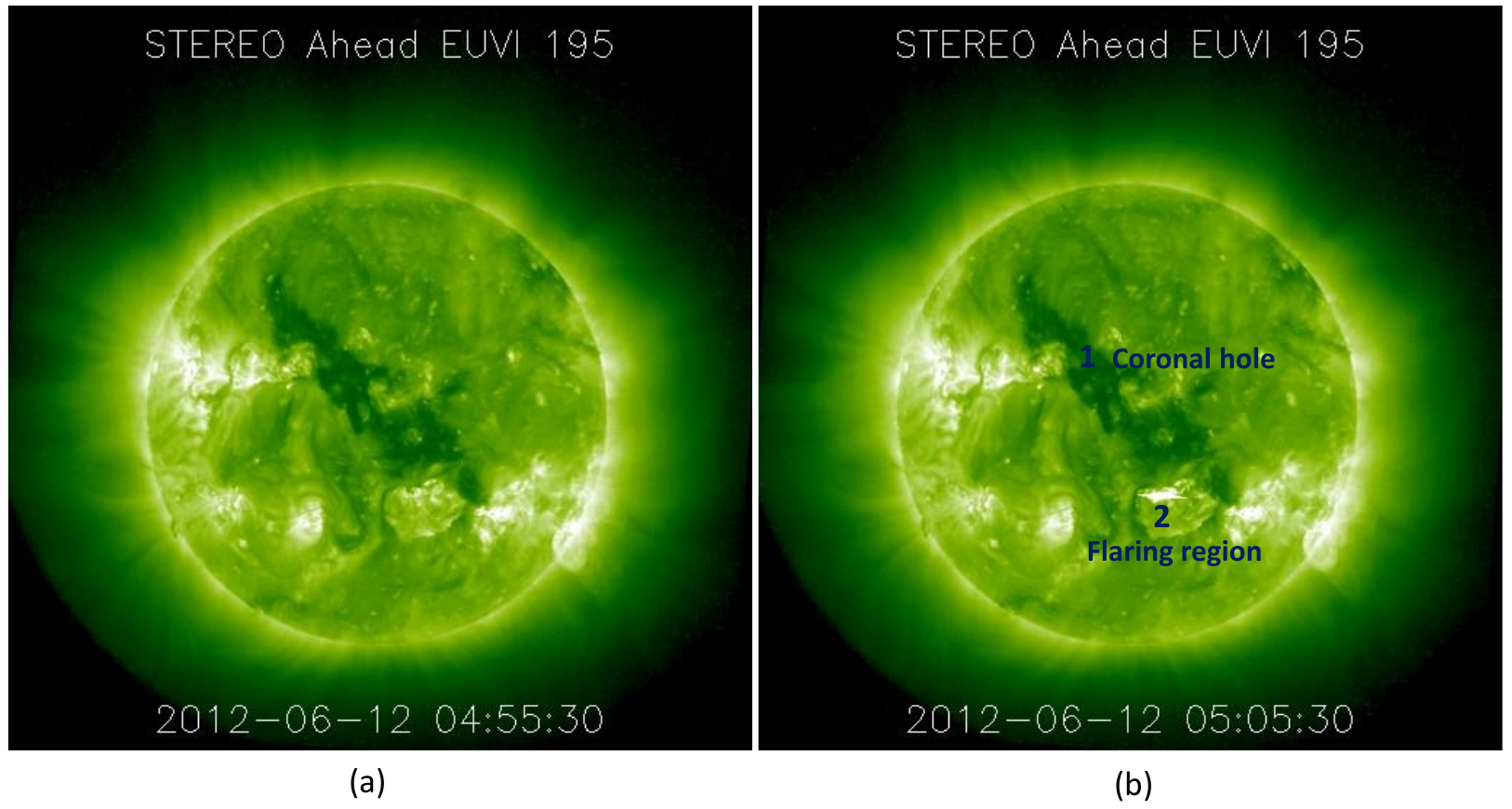}
\caption{Sun’s corona observed in 19.5 nm by the Extreme Ultraviolet Imager (EUVI) of the Sun Earth Connection Coronal and Heliospheric Investigation (SECCHI) instrument (e.g., \citealp{Wulser_et_al_2004}) on board STEREO-A at (a) 04:55:30 UT and  (b) 05:05:30 UT on June 12, 2012. The dark region (``1”) represents a coronal hole. The bright region (just above the number ``2”) is a flaring region. The location of this flare region is S15W126 \citep{Chertok_et_al_2015}.\label{fig:fig7}}
\end{figure}

\begin{figure}[h!]
\plotone{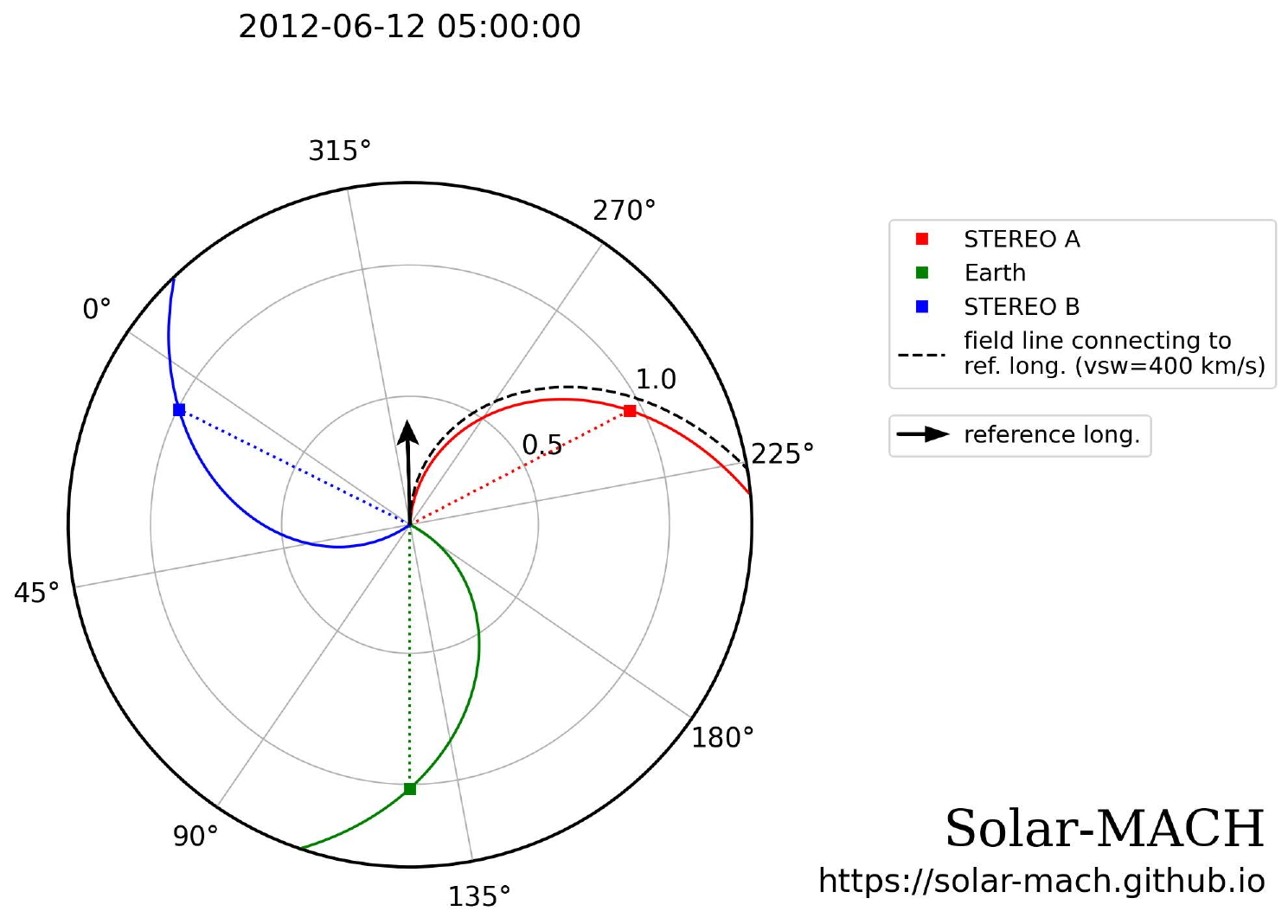}
\caption{Positions of STEREO-A (red), the Earth (green), and STEREO-B (blue) on Jun 12, 2012 at 05:00 UT. The colored solid spirals connecting these three objects to the Sun are calculated based on solar wind speed of 400 km/s. The solid arrow represents the location and direction of the solar flare mentioned in Figure \ref{fig:fig7}. The dashed spiral (black) is the magnetic field line originating from the flare location. This figure indicates that there is a possibility that STEREO-A was magnetically connected to the flaring region when SIR No. 12 passed the through it. Note that during this time, the Carrington longitudes of the Earth and STEREO-A were 127.2 and 244.6 respectively.\label{fig:fig8}}
\end{figure}

Going by the above arguments, SIR No. 12 stands out as an anomalous and special case. Despite having different FIP and m/q, suprathermal $^4$He, O, and Fe exhibit nearly identical spectral index (close to 1.5) in SIR No. 12. In a series of papers, \cite{Fisk_and_Gloeckler_2006, Fisk_and_Gloeckler_2008, Fisk_and_Gloeckler_2014} argued that compressional turbulences could accelerate particles when in steady states, these particles do equal amount of work on the turbulences. The source of free energy in this process is the pressure variations in the core particles. This free energy flows upwards in energy through suprathermal tails (energy cascading) that exhibit a power law index of 1.5 (see \citealp{Fisk_and_Gloeckler_2006} for more details). However, this condition is evidently not getting satisfied for suprathermal particles associated with other SIR events investigated in this work. \cite{Mason_et_al_2008a} and \cite{Filwett_et_al_2019} did not find any clustering of spectral indices near 1.5 value. Most interestingly, the ``pump mechanism", introduced by \cite{Fisk_and_Gloeckler_2008} and claimed to be responsible for 1.5 spectral index, is valid only for quiet solar wind i.e., devoid of any shocks or large scale compression regions. Since, we observe spectral index very close to 1.5 in $^4$He, O, and Fe spectra in connection with a SIR event (SIR No. 12), we can rule out the role of ``pump mechanism" in this case.  In the following part, we explore some other possibilities for 1.5 spectral index for different species. 

\begin{figure}[h!]
\plotone{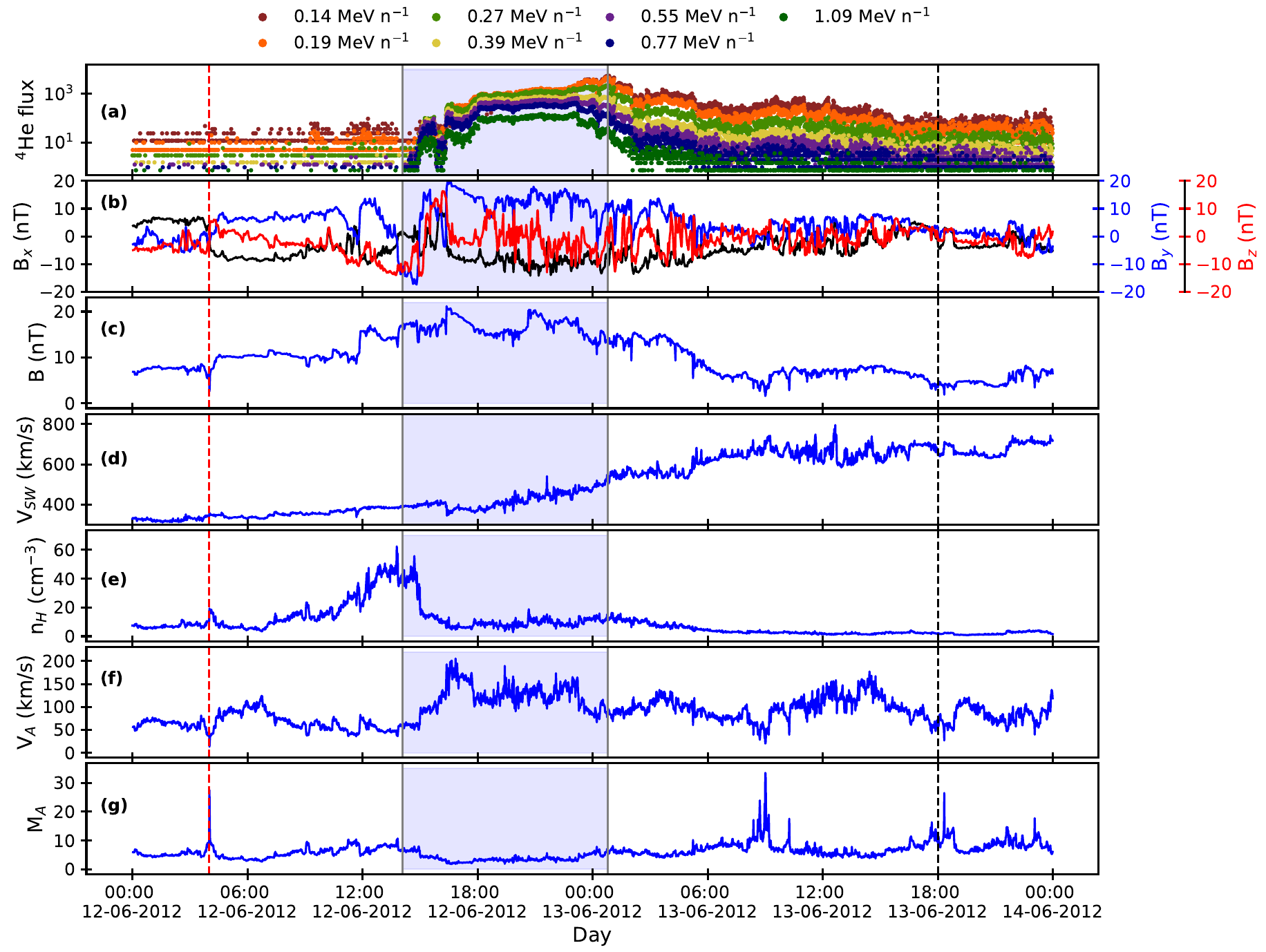}
\caption{Variations of (a) $^4$He fluxes corresponding to different energy channels, (b) components (B$_x$, B$_y$, and B$_z$) of interplanetary magnetic field (IMF) in RTN coordinate system, (c) magnitude of magnetic field (B), (d) bulk solar wind speed (V$_{SW}$), (e) proton number density (n$_H$), (f) Alfv\'en speed (V$_A$), and (g) Alfv\'enic Mach number (M$_A$) during SIR No. 12. The red and black dashed lines indicate the start and end time of the SIR as mentioned in Figure \ref{fig:fig1} and Table \ref{tab:table1}. The interval between the start of $^4$He flux enhancement and the time of peak 0.14 MeV n$^{-1}$ flux is shaded. \label{fig:fig9}}
\end{figure}
\begin{figure}[ht!]
\plotone{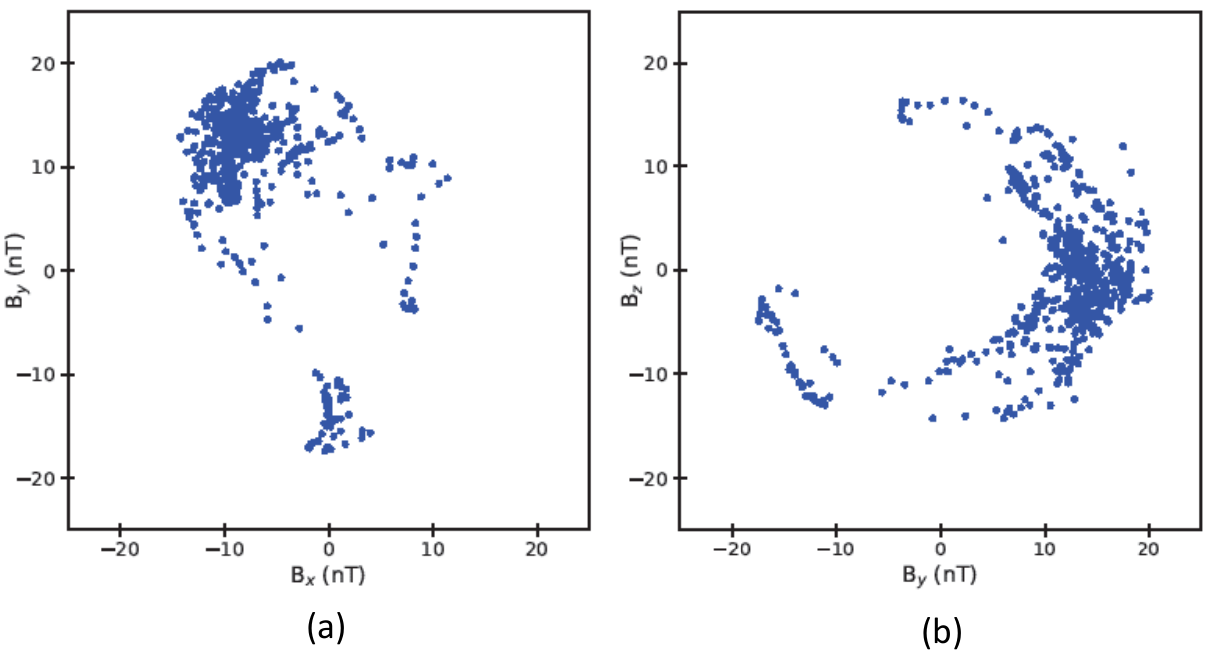}
\caption{Hodograms of IMF (a) B$_x$ – B$_y$ and (b) B$_y$ – B$_z$ during the shaded interval as shown in Figure \ref{fig:fig9}. The rotation in magnetic field components establishes the existence of magnetic islands during this interval. \label{fig:fig10}}
\end{figure}
In order to investigate SIR No. 12, we show Figure \ref{fig:fig7}. This figure shows the images of solar corona in 19.5 nm (extreme ultraviolet) light observed by Sun-Earth Connection Coronal and Heliospheric Investigation (SECCHI) instrument on board STEREO-A on June 12, 2012 at around (a) 04:55 UT and (b) 05:05 UT. One can see from Figure \ref{fig:fig7} that a large coronal hole was present around this time that extends from north-east to south-west mid-latitudes of the Sun and encompasses the equatorial region as well. The locations of STEREO-A, the Earth, and STEREO-B around the same time are shown in Figure \ref{fig:fig8}. During this time, the Carrington longitudes of the Earth and STEREO-A were 127.2 and 244.6 respectively. Therefore, this coronal hole was not visible from the Earth or STEREO-B. It is possible that SIR No. 12 (starts on Jun 12, 2012 at 04:00 UT) is generated due to fast solar wind stream emanating from this large coronal hole. Interestingly, SECCHI images also reveal onset of a flare at 05:05 UT. The location of the flaring region was at the edge of this coronal hole (as marked by ``2” in Figure 7). \cite{Chertok_et_al_2015} identified this flare event as a M4.6 class by analyzing length of the blooming streak associated with this flare. As can be seen from Figure 8, the magnetic field line connecting the flare location (S15W126, from \citealp{Chertok_et_al_2015}) propagates very close to STEREO-A.  Therefore, it is possible that the sharp enhancements in suprathermal $^4$He, O, and Fe fluxes occur due to an impulsive solar energetic particle (ISEP) event associated with the solar flare as shown in Figure \ref{fig:fig7}. \cite{Mason_et_al_2000, Mason_et_al_2002} showed by using data obtained from ULEIS and Solar Isotope Spectrometer (SIS) on board the ACE spacecraft that a particular class of $^3$He-rich ISEP events exhibit power laws in the heavy ion spectra (like $^3$He, $^4$He, O, and Fe) in the energy range 0.1 – 1.0 MeV n$^{-1}$. \cite{Mason_et_al_2002} reported an ISEP event observed by the ACE on Sep 09, 1998 where $<$ 1.5 MeV n$^{-1}$ $^3$He, $^4$He, O, and Fe exhibit similar variations in their spectra and the spectral indices lie in the range 1.15 – 1.38. Although role of the solar flare-associated ISEP event looks promising in producing identical spectral index, a counter evidence exists in the literature. Typical time delay between flare onset (measured by GOES X-ray fluxes) and ISEP detection closure to the Earth is around 1 hour (e.g., \citealp{Papaioannou_et_al_2023}). In case of SIR No. 12, this delay is around 9 hours. Even, in some ISEP events, dispersionless enhancement of different energy channels is not guaranteed (see \citealp{Mason_et_al_2000}). This essentially means that enhancement of any particle (for example, $^4$He) with a higher energy (say 1 MeV n$^{-1}$) starts earlier than a particle with lower energy (say 0.5 MeV n$^{-1}$). Taking a nominal Parker spiral path length of 1.2 au between the flaring region shown in Figure \ref{fig:fig7} and the location of STEREO-A, the time delay between a 0.5 MeV n$^{-1}$ and a 1 MeV n$^{-1}$ particle (for any particle because the energy unit is normalized by mass number) is calculated to be more than an hour. However, we do not find any noticeable time delay in $^4$He flux enhancements corresponding to 0.55 MeV n$^{-1}$ and 1.09 MeV n$^{-1}$ in SIR No. 12. Therefore, flux enhancements observed in SIR No. 12 do not appear to come directly from the flaring region. There may be another process at play in the IP medium, which is responsible for insignificant time delay and the identical spectral index of three different elements with different FIP and m/q. 

In this context, we believe that a more suitable mechanism that can explain this event would be acceleration of particles by merging and contraction of small-scale magnetic islands (SMIs) (\citealp{Cargill_et_al_2006, Drake_et_al_2006a, Drake_et_al_2006b, Bian_and_Kontar_2013, Zank_et_al_2014, Roux_et_al_2015, Khabarova_et_al_2015a} etc.). These SMIs can be induced in the vicinity of heliospheric current sheet (\citealp{Cartwright_and_Moldwin_2008, Bemporad_2008, Khabarova_et_al_2015a, Khabarova_et_al_2015b, Khabarova_et_al_2016} etc.) by different active processes in the solar corona (like solar flare, coronal mass ejections) or by various instabilities in the current sheets \citep{Eastwood_et_al_2002, Drake_et_al_2006a}. The typical length scale of such SMIs is 10$^6$ – 10$^7$ km \citep{Khabarova_et_al_2015a}. \cite{Zank_et_al_2014} showed analytically that particles accelerated due to merging and contraction of magnetic islands exhibit power law and the spectral indices depend on the Alfv\'enic Mach (M$_A$) number, which is independent of m/q of particles. At 1 au, a special case of this theory (M$_A$=7) predicts a velocity distribution spectrum that varies with V$^{-5}$ (V is the particle speed), which is equivalent to a differential directional flux spectrum $\sim$ E$^{-1.5}$ (E is the energy of the particle). To check the applicability of Zank et al’s theory in our observation of 1.5 spectral indices, we show Figure \ref{fig:fig9} and Figure \ref{fig:fig10}. In Figure \ref{fig:fig9}, we have plotted (a) $^4$He fluxes corresponding to different energy channels, (b) components (B$_x$, B$_y$, and B$_z$) of interplanetary magnetic field (IMF) in radial-tangential-normal (RTN) coordinate system, (c) magnitude of magnetic field (B), (d) bulk solar wind speed (V$_{SW}$), (e) proton number density (n$_H$), (f) Alfv\'en speed (V$_A$), and (g) Alfv\'enic Mach number (M$_A$) during SIR No. 12. It can be seen from Figure 9 (g) that M$_A$ varies in the range 5 – 10 several hours before the start of $^4$He flux enhancements. This is consistent with the proposition of \cite{Zank_et_al_2014}. More importantly, the very presence of magnetic islands are established from the hodograms (e.g., \citealp{Khabarova_et_al_2015a}) shown in Figure \ref{fig:fig10} (a) and (b), where rotation in the interplanetary magnetic field inside magnetic islands (shaded interval in Figure \ref{fig:fig9}) is evident. This kind of rotation is not observed in other SIR events. We also calculate typical length scale ($l$) by using Eq. 2 in \cite{Khabarova_et_al_2015b}, which reads $l = r\times (B/B_0)^{1/2}$ km, 
where r is distance in au, B = B(r) is the magnetic field, B$_0$=10,000 nT is the minimum coronal magnetic field provided by \cite{Jensen_and_Russell_2009}. By taking typical value of B =15 nT inside the shaded region in Figure \ref{fig:fig9}, we get $l \sim 6 \times 10^6$ km at 1 au. 

Therefore, it appears that the identical spectral indices (very close to 1.5) for $^4$He, O, and Fe associated with SIR No. 12 are caused due to local acceleration of these particles by SMIs, which are generated by the flare (shown in Figure \ref{fig:fig7}). Local acceleration of particles also supports the minimal modulation (no significant changes in the spectral indices) of these particles before detected by STEREO-A along with characteristic of source process.

\section{Summary}
This study investigates twenty shock-less stream interaction region (SIR) events observed by the STEREO-A spacecraft to understand better the energization process of suprathermal particles, especially those not linked to shocks. We analyze the differential directional flux of three distinct species ($^{4}$He, O, and Fe) across energy ranges from $0.14$ to $1.09$ MeV per nucleon and calculate their respective spectral indices for each event. Most of our events show significant variations in spectral indices for each of the three species. Additionally, we observe that the spectral index for each species varies over a wide range -- $^{4}$He has a range of $1.55$ to $4.08$, O showed variability between $1.49$ and $4.56$, and Fe ranged from $1.46$ to $4.04$. We ascribe these spectral index variations to the stochastic nature of their origins and to interplanetary modulation, which depends on the $m/q$ ratio.

In one of the twenty events, which we refer to as SIR No. 12, we observe that the spectral indices for all three species remain close to $1.5$. We found that this event is associated with an M4.6-class solar flare that occurred approximately nine hours before the observed flux enhancement, and this connection is traced along the Parker spiral. Further investigation of this event reveals that during this period, the magnetic field rotates in the $y-z$ plane, perpendicular to the equatorial plane ($x-y$ plane). Additionally, we note that the solar wind's Alfv\'enic Mach number ($M_A$) prior to the event remains within the range of $7-10$. Taking all of these pieces of information into account, we conclude that the suprathermal population is generated locally through the merging and contraction of small-scale magnetic islands, possibly formed as a result of the flare and situated both above and below the heliospheric current sheet. For this particular case, we calculate the size of the typical small-scale magnetic islands to be approximately $\sim 6 \times 10^6$ km.

\section{Acknowledgements}
We are grateful to the Principal Investigators and all the members of the SIT, PLASTIC, and MAG (on board STEREO-A) team for constructing these instruments and thereafter, generating and managing the data sets used in this work. We thank Mihir I. Desai, Janet Luhmann, Peter Schroeder, and Glenn M. Mason for clarifying certain aspects of SIT data used in this paper. We express our sincere gratitude to the Department of Space, Government of India for supporting this work. 

\section{Data availability}
The list of SIR events used in this work is available at \url{https://stereo-dev.epss.ucla.edu/media/SIRs.pdf}. Suprathermal flux data are obtained from the SIT/IMPACT, in-situ data from PLASTIC/IMPACT, and magnetic field data from MAG/IMPACT on board the STEREO-A are available at \url{https://cdaweb.gsfc.nasa.gov/index.html}. IP shock list observed by the STEREO – A is available at http://www.ipshocks.fi/database. EUVI/SECCHI images are available at \url{https://cdaw.gsfc.nasa.gov/}.

\bibliography{BD_paper_2_ApJ}{}

\begin{thebibliography}{}
\expandafter\ifx\csname natexlab\endcsname\relax\def\natexlab#1{#1}\fi
\providecommand{\url}[1]{\href{#1}{#1}}
\providecommand{\dodoi}[1]{doi:~\href{http://doi.org/#1}{\nolinkurl{#1}}}
\providecommand{\doeprint}[1]{\href{http://ascl.net/#1}{\nolinkurl{http://ascl.net/#1}}}
\providecommand{\doarXiv}[1]{\href{https://arxiv.org/abs/#1}{\nolinkurl{https://arxiv.org/abs/#1}}}

\bibitem[{Acu{\~n}a {et~al.}(2008)Acu{\~n}a, Curtis, Scheifele, Russell,
  Schroeder, Szabo, \& Luhmann}]{Acuna_et_al_2008}
Acu{\~n}a, M., Curtis, D., Scheifele, J., {et~al.} 2008, Space Science Reviews,
  136, 203

\bibitem[{Agresti \& Coull(1998)}]{Agresti_and_Coull_1998}
Agresti, A., \& Coull, B.~A. 1998, The American Statistician, 52, 119

\bibitem[{Allen {et~al.}(2019)Allen, Ho, \& Mason}]{Allen_et_al_2019}
Allen, R.~C., Ho, G.~C., \& Mason, G.~M. 2019, The Astrophysical Journal
  Letters, 883, L10

\bibitem[{Allen {et~al.}(2021)Allen, Mason, Ho, Rodriguez-Pacheco,
  Wimmer-Schweingruber, Andrews, Berger, Boden, Cernuda, Lara,
  {et~al.}}]{Allen_et_al_2021}
Allen, R.~C., Mason, G.~M., Ho, G.~C., {et~al.} 2021, Astronomy \&
  Astrophysics, 656, L2

\bibitem[{Belcher \& Davis~Jr(1971)}]{Belcher_and_Davis_1971}
Belcher, J., \& Davis~Jr, L. 1971, Journal of Geophysical Research, 76, 3534

\bibitem[{Bemporad(2008)}]{Bemporad_2008}
Bemporad, A. 2008, The Astrophysical Journal, 689, 572

\bibitem[{Bian \& Kontar(2013)}]{Bian_and_Kontar_2013}
Bian, N.~H., \& Kontar, E.~P. 2013, Physical review letters, 110, 151101

\bibitem[{Bu{\v{c}}{\'\i}k {et~al.}(2009)Bu{\v{c}}{\'\i}k, Mall, Korth, \&
  Mason}]{Buvcik_et_al_2009}
Bu{\v{c}}{\'\i}k, R., Mall, U., Korth, A., \& Mason, G. 2009in , Copernicus
  Publications G{\"o}ttingen, Germany, 3677--3690

\bibitem[{Cargill {et~al.}(2006)Cargill, Vlahos, Turkmani, Galsgaard, \&
  Isliker}]{Cargill_et_al_2006}
Cargill, P.~J., Vlahos, L., Turkmani, R., Galsgaard, K., \& Isliker, H. 2006,
  Solar Dynamics and Its Effects on the Heliosphere and Earth, 124, 249

\bibitem[{Cartwright \& Moldwin(2008)}]{Cartwright_and_Moldwin_2008}
Cartwright, M., \& Moldwin, M. 2008, Journal of Geophysical Research: Space
  Physics, 113

\bibitem[{Chertok {et~al.}(2015)Chertok, Belov, \&
  Grechnev}]{Chertok_et_al_2015}
Chertok, I., Belov, A., \& Grechnev, V. 2015, Solar Physics, 290, 1947

\bibitem[{Dalal {et~al.}(2022)Dalal, Chakrabarty, \&
  Srivastava}]{Dalal_et_al_2022}
Dalal, B., Chakrabarty, D., \& Srivastava, N. 2022, The Astrophysical Journal,
  938, 26

\bibitem[{Desai {et~al.}(1998)Desai, Marsden, Sanderson, Balogh, Forsyth, \&
  Gosling}]{Desai_et_al_1998}
Desai, M., Marsden, R., Sanderson, T., {et~al.} 1998, Journal of Geophysical
  Research: Space Physics, 103, 2003

\bibitem[{Desai {et~al.}(2004)Desai, Mason, Wiedenbeck, Cohen, Mazur, Dwyer,
  Gold, Krimigis, Hu, Smith, {et~al.}}]{Desai_et_al_2004}
Desai, M.~I., Mason, G.~M., Wiedenbeck, M.~E., {et~al.} 2004, The Astrophysical
  Journal, 611, 1156

\bibitem[{Drake {et~al.}(2006{\natexlab{a}})Drake, Swisdak, Che, \&
  Shay}]{Drake_et_al_2006a}
Drake, J., Swisdak, M., Che, H., \& Shay, M. 2006{\natexlab{a}}, Nature, 443,
  553

\bibitem[{Drake {et~al.}(2006{\natexlab{b}})Drake, Swisdak, Schoeffler, Rogers,
  \& Kobayashi}]{Drake_et_al_2006b}
Drake, J., Swisdak, M., Schoeffler, K., Rogers, B., \& Kobayashi, S.
  2006{\natexlab{b}}, Geophysical research letters, 33

\bibitem[{Eastwood {et~al.}(2002)Eastwood, Balogh, Dunlop, \&
  Smith}]{Eastwood_et_al_2002}
Eastwood, J., Balogh, A., Dunlop, M., \& Smith, C. 2002, Journal of Geophysical
  Research: Space Physics, 107, SSH

\bibitem[{Ebert {et~al.}(2012)Ebert, Dayeh, Desai, \& Mason}]{Ebert_et_al_2012}
Ebert, R., Dayeh, M., Desai, M., \& Mason, G. 2012, The Astrophysical Journal,
  749, 73

\bibitem[{Filwett {et~al.}(2019)Filwett, Desai, Ebert, \&
  Dayeh}]{Filwett_et_al_2019}
Filwett, R., Desai, M., Ebert, R., \& Dayeh, M. 2019, The Astrophysical
  Journal, 876, 88

\bibitem[{Fisk \& Gloeckler(2006)}]{Fisk_and_Gloeckler_2006}
Fisk, L., \& Gloeckler, G. 2006, The Astrophysical Journal Letters, 640, L79

\bibitem[{Fisk \& Gloeckler(2008)}]{Fisk_and_Gloeckler_2008}
---. 2008, The Astrophysical Journal, 686, 1466

\bibitem[{Fisk \& Gloeckler(2014)}]{Fisk_and_Gloeckler_2014}
---. 2014, Journal of Geophysical Research: Space Physics, 119, 8733

\bibitem[{Fisk \& Lee(1980)}]{Fisk_and_Lee_1980}
Fisk, L., \& Lee, M. 1980, The Astrophysical Journal, 237, 620

\bibitem[{Fisk \& Gloeckler(2007)}]{Fisk_and_Gloeckler_2007}
Fisk, L.~A., \& Gloeckler, G. 2007, Proceedings of the National Academy of
  Sciences, 104, 5749

\bibitem[{Galvin {et~al.}(2008)Galvin, Kistler, Popecki, Farrugia, Simunac,
  Ellis, M{\"o}bius, Lee, Boehm, Carroll, {et~al.}}]{Galvin_et_al_2008}
Galvin, A.~B., Kistler, L.~M., Popecki, M.~A., {et~al.} 2008, The STEREO
  Mission, 437

\bibitem[{Giacalone {et~al.}(2002)Giacalone, Jokipii, \&
  Kota}]{Giacalone_et_al_2002}
Giacalone, J., Jokipii, J.~R., \& Kota, J. 2002, The Astrophysical Journal,
  573, 845, \dodoi{10.1086/340660}

\bibitem[{Gloeckler(2003)}]{Gloeckler_2003}
Gloeckler, G. 2003in , American Institute of Physics, 583--588

\bibitem[{Gosling {et~al.}(1981)Gosling, Borrini, Asbridge, Bame, Feldman, \&
  Hansen}]{Gosling_et_al_1981}
Gosling, J., Borrini, G., Asbridge, J., {et~al.} 1981, Journal of Geophysical
  Research: Space Physics, 86, 5438

\bibitem[{Jensen \& Russell(2009)}]{Jensen_and_Russell_2009}
Jensen, E., \& Russell, C. 2009, Geophysical research letters, 36

\bibitem[{Jian {et~al.}(2019)Jian, Luhmann, Russell, \&
  Galvin}]{Jian_et_al_2019}
Jian, L., Luhmann, J., Russell, C., \& Galvin, A. 2019, Solar physics, 294, 1

\bibitem[{Jian {et~al.}(2013)Jian, Russell, Luhmann, Galvin, \&
  Simunac}]{Jian_et_al_2013}
Jian, L., Russell, C., Luhmann, J., Galvin, A., \& Simunac, K. 2013in ,
  American Institute of Physics, 191--194

\bibitem[{Jian {et~al.}(2006)Jian, Russell, Luhmann, \&
  Skoug}]{Jian_et_al_2006}
Jian, L., Russell, C., Luhmann, J., \& Skoug, R. 2006, Solar Physics, 239, 337

\bibitem[{Khabarova {et~al.}(2015{\natexlab{a}})Khabarova, Zank, Li, Le~Roux,
  Webb, Dosch, \& Malandraki}]{Khabarova_et_al_2015a}
Khabarova, O., Zank, G., Li, G., {et~al.} 2015{\natexlab{a}}, The Astrophysical
  Journal, 808, 181

\bibitem[{Khabarova {et~al.}(2015{\natexlab{b}})Khabarova, Zank, Li, le~Roux,
  Webb, Malandraki, \& Zharkova}]{Khabarova_et_al_2015b}
Khabarova, O.~V., Zank, G.~P., Li, G., {et~al.} 2015{\natexlab{b}}in , IOP
  Publishing, 012033

\bibitem[{Khabarova {et~al.}(2016)Khabarova, Zank, Li, Malandraki, le~Roux, \&
  Webb}]{Khabarova_et_al_2016}
Khabarova, O.~V., Zank, G.~P., Li, G., {et~al.} 2016, The Astrophysical
  Journal, 827, 122

\bibitem[{Le~Roux {et~al.}(2015)Le~Roux, Zank, Webb, \&
  Khabarova}]{Roux_et_al_2015}
Le~Roux, J., Zank, G., Webb, G., \& Khabarova, O. 2015, The Astrophysical
  Journal, 801, 112

\bibitem[{Luhmann {et~al.}(2008)Luhmann, Curtis, Schroeder, McCauley, Lin,
  Larson, Bale, Sauvaud, Aoustin, Mewaldt, {et~al.}}]{Luhmann_et_al_2008}
Luhmann, J., Curtis, D., Schroeder, P., {et~al.} 2008, in The STEREO Mission
  (Springer), 117--184

\bibitem[{Mason {et~al.}(2012)Mason, Desai, \& Li}]{Mason_et_al_2012}
Mason, G., Desai, M., \& Li, G. 2012, The Astrophysical Journal Letters, 748,
  L31

\bibitem[{Mason {et~al.}(2000)Mason, Dwyer, \& Mazur}]{Mason_et_al_2000}
Mason, G., Dwyer, J., \& Mazur, J. 2000, The Astrophysical Journal, 545, L157

\bibitem[{Mason {et~al.}(2008{\natexlab{a}})Mason, Korth, Walpole, Desai,
  Von~Rosenvinge, \& Shuman}]{Mason_et_al_2008b}
Mason, G., Korth, A., Walpole, P., {et~al.} 2008{\natexlab{a}}, Space science
  reviews, 136, 257

\bibitem[{Mason {et~al.}(2002)Mason, Wiedenbeck, Miller, Mazur, Christian,
  Cohen, Cummings, Dwyer, Gold, Krimigis, {et~al.}}]{Mason_et_al_2002}
Mason, G.~M., Wiedenbeck, M.~E., Miller, J.~A., {et~al.} 2002, The
  Astrophysical Journal, 574, 1039

\bibitem[{Mason {et~al.}(2008{\natexlab{b}})Mason, Leske, Desai, Cohen, Dwyer,
  Mazur, Mewaldt, Gold, \& Krimigis}]{Mason_et_al_2008a}
Mason, G.~M., Leske, R.~A., Desai, M.~I., {et~al.} 2008{\natexlab{b}}, The
  Astrophysical Journal, 678, 1458

\bibitem[{Papaioannou {et~al.}(2023)Papaioannou, Herbst, Ramm, Cliver, Lario,
  \& Veronig}]{Papaioannou_et_al_2023}
Papaioannou, A., Herbst, K., Ramm, T., {et~al.} 2023, Astronomy \&
  Astrophysics, 671, A66

\bibitem[{Richardson(2018)}]{Richardson_2018}
Richardson, I.~G. 2018, Living reviews in solar physics, 15, 1

\bibitem[{Schwadron {et~al.}(2010)Schwadron, Dayeh, Desai, Fahr, Jokipii, \&
  Lee}]{Schwadron_et_al_2010}
Schwadron, N.~A., Dayeh, M., Desai, M., {et~al.} 2010, The Astrophysical
  Journal, 713, 1386

\bibitem[{Tsurutani {et~al.}(1982)Tsurutani, Smith, Pyle, \&
  Simpson}]{Tsurutani_et_al_1982}
Tsurutani, B., Smith, E., Pyle, K., \& Simpson, J. 1982, Journal of Geophysical
  Research: Space Physics, 87, 7389

\bibitem[{W{\"u}lser {et~al.}(2004)W{\"u}lser, Lemen, Tarbell, Wolfson, Cannon,
  Carpenter, Duncan, Gradwohl, Meyer, Moore, {et~al.}}]{Wulser_et_al_2004}
W{\"u}lser, J.-P., Lemen, J.~R., Tarbell, T.~D., {et~al.} 2004, in Telescopes
  and instrumentation for solar astrophysics, Vol. 5171, SPIE, 111--122

\bibitem[{Zank {et~al.}(2014)Zank, Le~Roux, Webb, Dosch, \&
  Khabarova}]{Zank_et_al_2014}
Zank, G.~l., Le~Roux, J., Webb, G., Dosch, A., \& Khabarova, O. 2014, The
  Astrophysical Journal, 797, 28

\end{thebibliography}
\bibliographystyle{aasjournal}

\end{document}